\documentclass[12pt]{article}
\usepackage[dvips,
hyperfigures=true,
]{hyperref}

\usepackage{graphicx}
\usepackage{amssymb}
\title{Fixed point theorem for simple quantum strategies in quantum market games}
\author{Edward W\mbox{.} Piotrowski\\Institute of Theoretical Physics,
University of Bia\l ystok,\\ Lipowa 41, PL-15424 Bia\l ystok,
Poland\\
\href{http://alpha.uwb.edu.pl/ep/sj}{http://alpha.uwb.edu.pl/ep/sj}}
\begin{document}
\maketitle
\begin{abstract}
A simple but nontrivial class of the quantum strategies
 in buying-selling games is presented.
 The player moves are a rational buying and an
unconditional selling. The possibility of gaining extremal profits
in such the games is considered. {\em The  entangled merchants}\/
hypothesis is proposed.
\end{abstract}
\section{Projective geometry approach to profits}
Half thousand years ago Fra Luca Pacioli, a teacher of Leonardo da
Vinci set standards of Venice accounting (i.e\mbox{.} the double
accounting) and for the theory of perspective. In the author
opinion this is not accidental  because projective geometry
approach forms a natural language of description of the market reality.
Let us consider the simplest possible market event of exchanging
two goods which we would call the asset and the money and denote
them by $\Theta $ and $\$ $, respectively. Let $V_{\Theta}$ and $V_{\$
}$ denote some given amounts of the asset and the money,
respectively. If the assets are exchanged in the proportion
$V_{\$}\negthinspace:\negthinspace V_{\Theta}$ then in the context of
selling processes we call the logarithmic quotation for the asset
$\Theta$
\begin{equation}
p:=\ln \left( V_{\$} \right) - \ln \left( V_{\Theta} \right)
\end{equation}
 {\em the demand profit}\/. Respectively, in context of buying, we call the number
\begin{equation}
q:= \ln \left( V_{\Theta} \right)-\ln \left( V_{\$} \right)
\end{equation}
 {\em the supply profit}\/.
From projective geometry point of view
 any market can be represented in the $N$\/-dimensional real projective space, $\Re
P^{N}$ that is $(N\negthinspace+\negthinspace1)$-dimensional vector space $\Re ^{N+1}$ (one
real coordinate for each asset) subject to the equivalence
relation $v\negthinspace\sim\negthinspace\lambda v$ for $v\in \Re ^{N+1}$
and $\lambda\neq0$.
For example we identify all portfolios having assets in the same proportions.
In this context separate profits $p$ or $q$, gained by selling or buying
respectively, are not invariant (coordinate free). The profit $p+q$ gained
during the whole buying-selling cycle is given by the logarithm of an appropriate
anharmonic (cross) ratio \cite{couranteng}, and is an
invariant (e.g. its numerical value does not depend on units
chosen to measure the assets). The anharmonic ratio for four
points lying on a given line, $A,B,C,D$ is the double ratio
of lengths of segments $\frac{AC}{AB}\negthinspace:\negthinspace\frac{DC}{DB}$ and is
denoted by $[A,B,C,D]$. In our case the anharmonic ratio  in
question, $[\Theta , U_{q}, U_{p} ,\$]$, concerns the pair of points:
\begin{equation}
 U_{q}:=\left( \upsilon \,
\mathrm{e}^{q},\upsilon ,\ldots \right) , \ U_{p}:=
\left( w ,w \, \mathrm{e}^{p},\ldots \right)
\label{defffi}
\end{equation}
 and the pair $\Theta$,
$\$ $. The last pair results from the crossing of the
hypersurfaces $\overline{\Theta}$ and $\overline{\$}$
corresponding to the portfolios consisting of only one asset
$\Theta $ or $\$ $, respectively and the line $U_{q}U_{p}$. The dots represent other
coordinates (not necessary equal for both points). The line
connecting $U_{q}$ and $ U_{p}$
may be represented by one-parameter family of vectors
$u(\lambda )$ with $\mu$\/-coordinates given as:
\begin{equation}
u_{\mu}\left( \lambda\right) := \lambda\left( U_{q} \right) _{\mu} +
\left( 1-\lambda \right)  \left(
U_{p} \right) _{\mu} \,.
\end{equation}
 This implies
that the values of $\lambda $ corresponding to the points $\Theta
$ and $\$ $ are given by the conditions:
$
 u_{0}\left( \lambda
_{\$}  \right) =\lambda _{\$} \left( U _{q}
\right) _{0} + \left( 1-\lambda _{\$} \right)  \left( U _{p} \right) _{0}
=0$, $
u_{1}\left( \lambda _{\Theta}  \right) =\lambda _{ \Theta } \left( U
_{q} \right) _{1} + \left( 1-\lambda _{ \Theta}
\right)  \left( U _{p } \right) _{1} =0
$.
 Substitution of  Equation $(\ref{defffi})$ leads to
 \begin{equation}
\lambda _{\$} =\frac{w}{w-\upsilon}
\ \, \mbox{and} \ \,
\lambda
_{\Theta}= \frac{w}{w -\upsilon \,\mathrm{e}^{-(p+q)} }\,.
\end{equation}
The calculation of the logarithm of the cross ratio $[\Theta ,
U_{q}, U_{p\,}, \$]$ on the line
$U_{q} U_{p}$ gives
\begin{equation}
\ln \left[\Theta , U_q,
U_p ,\$ \right] = \ln \left[
\frac{w}{w-\upsilon
\mathrm{e}^{-(p+q)} },1,0, \frac{w}{w-\upsilon} \right]
=  \ln \frac{\upsilon\,w\,\mathrm{e}^{p+q}}{\upsilon\,w } =p+q\,
.\,\,
\end{equation}
This quantity for successive transactions is the unique (up to
some factor) additive invariant of projective transformations.

\section{Quantum market games}
Traders usually employ intuitive strategies that if perceived as
objective "being" are not easily describable in a quantitative
way. But these relations form natural objects of interests of
accountants and econometricians that deal with market reality.
This situation has an analog in physical phenomena. In the Ithaca
interpretation of quantum mechanics {\em"correlations have
physical reality; that which they correlate does not"}\/
\cite{mermin}. Besides, only quantum theory allows to consider
self-conscious entities \cite{albert}. In the series of articles
posed by the author with Jan S\l adkowski\footnote{full texts of
these papers are available at the web site {\em
http://alpha.uwb.edu.pl/ep/sj \/}} the quantum theory of games
\cite{meyer,eisert,flitney} is generalized on the
infinite-dimensional Hilbert space and used to description of some
fundamental market phenomena. These models give  many interesting
results, unattainable in the framework of classical theories. For
example, quantum theory predicts the property of undividity of
attention
 of traders (no cloning theorem), the sudden and violent changes of prices
 can be explained by the quantum Zeno phenomenon. The theory unifies also
 the English auction with the Vickrey's one attenuating the motivation
 properties of the latter. There are apparent analogies with quantum
 thermodynamics that allow to interpret market equilibrium  as a state
 with vanishing financial risk flow.
In this article the author present another amazing property of
quantum market games\footnote{see also J\mbox{.} S\l adkowski,
"Giffen paradoxes in quantum market games", in the current issue}.
In this formalism, each player
modify his strategy $\psi\negthinspace\in\negthinspace H$ by
acting on the Hilbert space $H$ with tactics ${\mathcal
U}\negthinspace:\negthinspace H
\negthinspace\rightarrow\negthinspace H$. The strategies can
be interpreted as superpositions of trading decisions. For a
trader they form the "quantum board". The tactics $\mathcal U$ are
linear, usually unitary operators, and the strategy functions
$\psi(p)\negthinspace\in\negthinspace L^2$ have integrable square
modulus. The French method of presenting demans/supply curves is
based on the assumption that the demand is a function of prices
and is usually referred to as the Cournot convention. In this way
Born interpretation of the product
$\overline{\psi(p)}\,\psi(p)$ lead us to define supply curve as
the cumulative distribution function $\int_{-\infty}^p
\overline{\psi(p')}\,\psi(p')\,dp'$. The player buying quantum
tactics $\psi(q)$ are  Fourier transforms of his tactics $\psi(p)$
in selling processes and, consequently, the appropriate
observables $\mathcal{Q}$ and $\mathcal{P}$ are canonical ones,
$[\mathcal{P},\mathcal{Q}]=i\hslash_E$, where the economic
dimensionless constant $\hslash_E$ describes the minimal
inclination of the player to risk (by analogy with quantum
harmonic oscillator, see below). One strategy works in two
different occasions. This is a property that does not occur in any
classical theory.

\section{Risk profile of player strategies}
We can describe player strategy independently of representation of this
strategy because the Fourier transformation commutes with the risk operator
$\frac{1}{2m}\,\mathcal{P}^2+\frac{m}{2}\,\mathcal{Q}^2$.
The parameter
$m>0$ measures the risk asymmetry between
buying and selling positions.
The pure strategies that are consistent with the low of
supply (or low of demand)
are represented, following Hudson theorem \cite{piotrslad3}, by gaussian functions.
Besides the Gibbs mixed strategies represent equilibrium markets
 leads to the same results as some gaussian strategies with the
  modified value of the constant $\hslash_E$.
For this reason in typical market games the considerations
of models with gaussian strategies
$\psi_{a,m}(p)\negthinspace:=
\negthinspace\frac{1}{\sqrt[4]{2\pi m}}\,
\mathrm{e}^{-\frac{(p-a)^2}{4 m}}$ will be good enough. When the dispersion $m$ of
the demand profit tends to $0$ the distribution function $\overline{\psi_{a,m}(p)}\,\psi_{a,m}(p)$
tends to Dirac strategy $\delta(p-a)$. It means that the player fixed
{\em withdrawal price}\/ $\mathrm{e}^a$ below which
she or he gives the selling up.

\section{The simplest quantum market game}
\begin{figure}[h]
\begin{center}
\includegraphics[height=4.25cm, width=8.25cm]{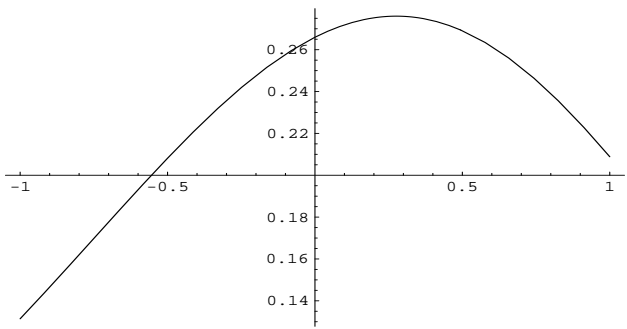}
\end{center}
\caption{Profit intensity for Dirac strategies in units of the opponent strategy dispersion $m^{-1}$.}
\label{hballfiij}
\end{figure}

The consequences of games between two Dirac strategies are trivial
and they are a special limit case of the class of
games analyzed below. Therefore we consider the games of Alice Dirac strategy
$\delta(p_{Alice}-a_{Alice})$ versus the gaussian strategy of Rest
of World (RW). Let us suppose that the RW player proposes price of
the asset $\Theta$ and Alice accepts or rejects the proposal because
this is the most attractive position for Alice. We have the following
rules of this game:
\begin{itemize}
\item RW proposes with probability
$f ( q_{RW}\negthinspace)\, dq_{RW}:=\sqrt{\frac{m}{2\pi}}\,
\mathrm{e}^{-\frac{m q_{\tiny RW}^2}{2}}\,dq_{RW}$ the price $\mathrm{e}^{-q_{RW}}$, and
\item Alice:
\begin{itemize}
\item sells the asset if $a_{Alice}+q_{RW}>0$, or
\item gives up if $a_{Alice}+q_{RW}\leq0$\/.
\end{itemize}
\end{itemize}
Alice always buys the asset in this game because the uncertainty
(dispersion) of stochastic variable $q_{Alice}$ for Dirac strategy
$\delta(p_{Alice}-a_{Alice})$ is infinite. We assume rational
behavior of Alice and therefore we search for her strategy
(i.e\mbox{.} value $a_{max}$ of variable $a_{Alice}$) which
maximize {\em the intensity}\/
 of her profit \cite{piotrslad2}:
\begin{equation}
\rho ( a_{Alice}):=
\frac{E(p_{RW}+q_{RW}\negthinspace)}{E(\tau)}=\frac{\int\limits
^{\infty}_{ a_{Alice}}q_{RW}f( q_{RW}\negthinspace)\, dq_{RW}}{1+
 \int\limits^{\infty}_{ a_{Alice}}f( q_{RW}\negthinspace)\, dq_{RW}} \,,
\end{equation}
where $E(p_{RW}+q_{RW}\negthinspace)$ is the average Alice profit in one
selling-buying cycle and
 $E(\tau)$ is the expected duration of the whole selling-buying cycle in this game.
The profit intensity function $\rho(x)$  in the vicinity of extremum
is drown in Figure \ref{hballfiij}.
\section{The fixed point theorem}
The maximal value of the function $\rho$\/, $a_{max}$, lies at a
fixed point of $\rho$, that is it fulfills the condition $\rho \left(
a_{max}\right) = a_{max}$. Such a fixed point $a_{max}$ exists, attracts
on all domain and
$a_{max} =0.27603\,m^{-1}$. When Rest of World play the
market game with strategy
 that has unknown dispersion $m^{-1}$ the above theorem proved in
 \cite{piotrslad2} implies
natural Alice tactics for maximization her intensity of profit. In
$n\negthinspace+\negthinspace1$-st
step of the game Alice shift her strategy to the point
\begin{equation}
a_{n+1}=\frac{\sum\limits_{k=1}^{n}(p_{RW}+q_{RW})_k}{\sum\limits_{k=1}^{n}\tau_k}\,.
\end{equation}
 One can easily reverse the buying
and selling strategies \cite{piotrslad2}.

When we modify rules of the game, and now Alice will be playing with
two entangled quantum strategies (one for buying and one for
selling), she will have another benefit of quanta. For the
expected intensity of Alice profit from one of her strategy we obtain
the equivalent of the above presented theorem again, but with greater
value of maximum equal $0.30211\,m^{-1}$ in this case. The details
will be presented elsewhere. In context of this result it is natural
to put forward the hypothesis that trading with bigger number of
entangled quantum strategies give us opportunities of obtaining
greater intensities of profits. Then in the future we will meet the
inevitable complication of quantum connections on real markets.
\bibliographystyle{amsalpha}

\end{document}